\documentstyle[twoside,fleqn,espcrc2]{article}
\input epsf

\newcommand{\AmS}{{\protect\the\textfont2
  A\kern-.1667em\lower.5ex\hbox{M}\kern-.125emS}}

\title{Finite Temperature Quark Confinement via Chromomagnetic Fields} 

\author{Michael Ogilvie\address{Department of Physics,
        Washington University, \\ 
        Campus Box 1105, St. Louis, MO, USA}%
\thanks{
This work was supported by the U.S. Department of Energy under grant number
DE-FG02-91-ER40628.}
        }
       
\begin{document}

\begin{abstract}
A natural mechanism for finite temperature quark confinement arises
via the coupling of the adjoint Polyakov loop to the chromomagnetic field.
Lattice simulations and analytical results both support this hypothesis.
Finite temperature SU(3) lattice
simulations show that a large external coupling to the chromomagnetic field
restores confinement at temperatures above the normal deconfining temperature.
A one-loop calculation of the effective potential for SU(2) gluons in a
background field shows that a constant chromomagnetic field can
drive the Polyakov loop to confining behavior, and the Polyakov loop can
in turn remove the well-known tachyonic mode associated with 
gluons in an external chromomagnetic field.
For abelian background fields, tachyonic modes
are necessary for confinement at one loop.

\end{abstract}

\maketitle

\section{INTRODUCTION}
The utility of the Polyakov loop ${\cal P}$ makes
finite temperature lattice gauge theory a natural place to study
confinement. A key issue is the mechanism which drives the Polyakov
loop expectation value to zero in the confined phase. 
In section 2, it is shown that a large external field coupled to
$F_{\mu\nu}$ restores confinement above the deconfinement temperature,
explicitly demonstrating a connection between the chromomagnetic field
and confinement. In section 3, analytical results for $SU(2)$ give a 
form for the coupling between ${\cal P}$ and $F_{\mu\nu}$.

\section{RESTORATION OF CONFINEMENT ABOVE $T_C$ BY AN APPLIED EXTERNAL FIELD}

An external field $J_{\mu\nu}$ can be coupled to the gauge field 
$F_{\mu\nu}$ by adding to the lattice action a term of the form
\begin{equation}
\sum_{x} \sum_{\mu > \nu} \sum_{a} {1 \over N}
\left[ J_{\mu\nu}^a(x) F_{\mu\nu}^a(x) \right]
\end{equation}
where the sums are over lattice sites, directions in space-time and
directions in the Lie algebra of the gauge group.
For the lattice form of the gauge potential, the simplest definition
is used:
\begin{equation}
F_{\mu\nu}(x) = {1 \over 2 i} \left[ U_{\mu\nu}(x) - U_{\mu\nu}^{+}(x)
\right].
\end{equation}
For lattice simulations, it is most convenient to not fix the gauge,
as would be necessary in the continuum. As a consequence, the partition
function $Z$ satifies
\begin{equation}
Z\left[ g(x) J_{\mu\nu}(x) g^{+}(x) \right] = Z\left[ J_{\mu\nu}(x) \right].
\end{equation}
For the case we consider where $J$ is non-zero in a single hyperplane,
this property of $Z$ under local gauge transformations implies that
$Z$ depends only on the eigenvalues of $J$.

Figure 1 shows the results of simulations of 
$SU(3)$ lattice gauge theory on a $16^3 \times 4 $ lattice
at $\beta = 6.0$, 
plotting the Polyakov loop $P$ against source $J_{12}^3$ and $J_{34}^3$.
The superscript $3$ indicates that the source is in the
$\lambda_3$ direction in the gauge group, while the subscript
indicates sources coupled to $F_{\mu\nu}$ in the $12$ and
$34$ direction, thus coupling to the real chromomagnetic field and
the imaginary chromoelectric field, respectively.
With $J=0$, $\beta=6.0$ at $N_t = 4$ is well into the deconfined
phase of finite temperature $SU(3)$.
The phase transition to the confined phase for 
sufficiently large $J$ is obvious in the figure.
Examination of time series and histograms indicate that the
transition observed in Figure 1 is most likely first order.
A comparison of $J_{12}^3$ and $J_{12}^8$ shows no evidence for
asymmetry in the gauge group: $\lambda_3$ and $\lambda_8$ appear
equivalent.
A small difference is seen between $J_{12}^3$
and $J_{34}^3$,
consistent with the more direct connection of the temporal plaquettes
to Polyakov loops via the link elements $U_4$.

\vspace{-0.3in}
\begin{figure}[htb]
\epsfxsize=75mm \epsfbox{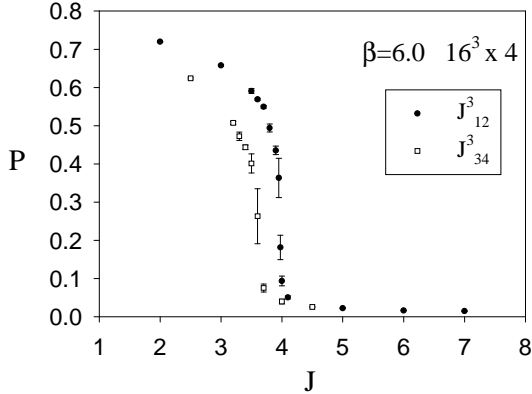}
\vspace{-0.3in}
\caption{$Tr_F ( {\cal P} )$ versus source $J$.}
\label{fig:fig1}
\end{figure}
\vspace{-0.3in}
The restoration of symmetry by a sufficiently strong external field
coupled to the chromomagnetic field is reminiscent of the restoration
of the normal phase from the superconducting phase by a strong
magnetic field.
It is natural from this point of view to consider
the system in a dimensionally reduced form, in which the Polyakov
loop play the role of a Higgs field in the adjoint representation,
coupled to a three-dimensional gauge field.

A complimentary approach is to view the effect of the external
field as changing the effective coupling constant. For the simple case
considered here, it is possible to find an alternative form for the
action by explicitly integrating over all local gauge transformations.
The leading term in $J_{\mu\nu}$ is proportional to
$ Tr \left[ J_{\mu\nu}^2 \right] Tr \left[ F_{\mu\nu}^2 \right] $,
with a negative sign, indicating a reduction in the effective value
of the gauge coupling $\beta$.

\section{QUARK CONFINEMENT BY CONSTANT FIELDS}

One plausible scenario for confinement
is that the coupling between the local
gauge field $F_{\mu\nu}$ and the adjoint Polyakov loop produces an effective
action which leads to two different phases \cite{MeOgGluon}.
In the
low temperature phase, there is a magnetic condensate and the Polyakov loop
indicates confinement; in the high temperature phase the magnetic condensate
vanishes, the Polyakov loop indicates deconfinement, and
the contribution of the thermal gauge boson gas dominates
the free energy.
This occurs because finite temperature effects
naturally couple the local gauge field to 
Polyakov loops\cite{MeOgI,MeOgII,MeOgIII,MeOgIV}.

For $SU(N)$,
the fundamental and adjoint representation Polyakov loops are related by
\begin{eqnarray}
Tr_A ( {\cal P} ) = | Tr_F ( {\cal P} ) |^2 - 1.
\end{eqnarray}
Clearly, when $Tr_A ( {\cal P} )$ assumes its minimum value of $-1$,
$Tr_F ( {\cal P} )$ assumes the value $0$.
Thus, one way to produce confinement in the low temperature phase is
for the free energy to be minimized by minimizing the expected value
of the trace of the adjoint Polyakov loop.

The case of a constant background chromagnetic field in $SU(2)$
illustrates this possibility.
The color magnetic field
and the Polyakov loop are taken 
to be simultaneously diagonal, and 
the color magnetic field $H$ points in the $x_3$ direction.
The Polyakov loop is specified by a constant $A_0$ field,
given in the adjoint representation by
$A_0 = \phi \tau_3 / {2 \beta}$.
The trace of the Polyakov loop is then given by
$Tr_F ( {\cal P} ) = 2 \cos (\phi / 2 )$
in the fundamental representation and by
$Tr_A ( {\cal P} ) = 1 + 2 \cos (\phi )$
in the adjoint representation.
The external magnetic field we take to have the form
$A_2 = H x_1 \tau_3 / 2$
which gives rise to a chromomagnetic field
$ F_{12} = H \tau_3 / 2 $.

The one-loop contribution to the free energy
has the usual form \cite{NiOl,NiSa,vanBaal} of a sum over the logarithms
of modes. The crucial contributions to this mode sum comes from
modes of the form
\begin{eqnarray}
\left[ \left( \omega_n - {\phi \over \beta} \right)^2 +
2 H \left( m + {1 \over 2} \pm 1 \right)  + k_3^2 \right]
\label{e2.1}
\end{eqnarray}
where the $\omega_n = 2 \pi n / \beta$ are the usual Matsubara frequencies, and
where the terms $2 H (m + 1/2 \pm 1)$ are the allowed Landau levels of the
gauge field in a background chromomagnetic field.

When $ \phi = 0$,
the $n = 0$ and $m = 0$ modes give rise to tachyonic modes
for $k_3$ sufficiently small;
these in turn give rise to an imaginary part in the free energy \cite{NiOl}.
These same modes will give a strictly real
factor to the determinant provided
\begin{eqnarray}
\beta \sqrt{H} < \phi < 2 \pi -  \beta \sqrt{H}.
\label{e2.3}
\end{eqnarray}

The renormalized effective potential has real component
\begin{eqnarray}
V_R &=& {11  H^2 \over 48 {\pi}^2 } \,
\ln \left( {H \over {\mu}_0^2} \right) 
- { 2 \pi^2 \over 90 \beta^4 }
- { ( H )^{3/2} \over {\pi}^2 \beta} \,\cdot
\nonumber
\end{eqnarray}
\begin{eqnarray}
\quad \sum_{n = 1}^{\infty} \,
{\cos (n \phi) \over n}
\bigg[ K_1 (n \beta \sqrt{H} )
- {\pi \over 2} Y_1 (n \beta \sqrt{H} )
\nonumber
\end{eqnarray}
\begin{eqnarray}
\quad + 2 \sum_{m = 0}^{\infty} \,
\sqrt{2 m + 3} K_1 [n \beta \sqrt{(2 m + 3)H}\, ]
\bigg]
\label{e2.11}
\end{eqnarray}
and imaginary component
\begin{eqnarray}
V_I = - { H^2 \over 8 \pi} - {H^{3/2} \over 2 \pi \beta} \,
\sum_{n = 1}^{\infty} \, {\cos (n \phi ) \over n} \,
J_1 ( n \beta \sqrt{H} ).
\label{e2.12}
\end{eqnarray}

At low
temperatures, minimization of $Re(V)$ leads to $\phi=\pi$ being
preferred. This is shown in Figure 2, which plots the real part of the
effective potential versus $H/\mu^2$ at $T/\mu = 0.25$
for both $\phi=0$ and $\phi=\pi$.
Unfortunately,
examination of $V_I$ shows that the lowest minima is not stable,
lying just to the right of the stable region, so that this
background field configuration remains unstable.
The confining solution $(\phi = \pi)$ 
is preferred over the perturbative vacuum
only for sufficiently low temperatures.
The tachyonic mode is responsible for $\phi=\pi$ being favored at low
temperature, and appears in the expression for $V_R$ as the $Y_1$ term.
Analysis shows that, at one loop for an arbitrary abelian background field, 
tachyonic modes must occur when $\phi=0$ if $\phi=\pi$ is to
be favored at low temperature. If there are no tachonic modes,
$\phi=0$ is always favored.

\begin{figure}[htb]
\epsfxsize=75mm \epsfbox{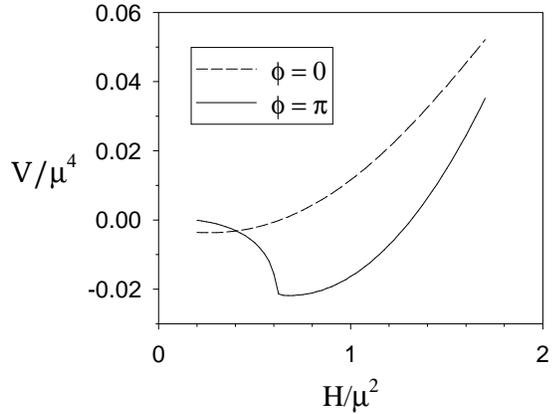}
\vspace{-0.3in}
\caption{$V/\mu^4$ at $T/\mu = 0.25$ for $\phi = 0$ and $\phi = \pi$.}
\label{fig:fig2}
\end{figure}

\section{CONCLUSIONS}

There is strong evidence from simulations and perturbation theory for the 
relevance of the chromomagnetic field in confinement, but we are still
far from builing a realistic model of the QCD vacuum.
Directions for future research along the
lines discussed here include
simulations and analytical work
to determine the effect of a quenched
random field $J$ and
to explore the behavior of instantons, monopoles and vortices
in an external field.

\end{document}